\documentclass[twocolumn,superscriptaddress,floatfix,prb]{revtex4-1}

\usepackage{graphicx,xcolor}

\usepackage{amssymb,amsmath}
\usepackage[mathscr]{eucal}
\usepackage{textcomp}

\newcommand{\blue}[1]{\textcolor{black}{#1}}

\begin{document}

\title{Kondo effect in binuclear metal-organic complexes with weakly interacting spins}

\author{L.\ Zhang}
\affiliation{Institute of Nanotechnology, Karlsruhe Institute of Technology (KIT), Germany}
\affiliation{Physikalisches Institut, Karlsruhe Institute of Technology (KIT), Germany}

\author{A.\ Bagrets}
\affiliation{Institute of Nanotechnology, Karlsruhe Institute of Technology (KIT), Germany}

\author{D.\ Xenioti}
\affiliation{Institut de Physique et Chimie des Mat{\'e}riaux de Strasbourg (IPCMS), Strasbourg, France}
\affiliation{Institute of Nanotechnology, Karlsruhe Institute of Technology (KIT), Germany}

\author{R.\ Koryt\'{a}r}
\affiliation{Institute of Nanotechnology, Karlsruhe Institute of Technology (KIT), Germany}

\author{M.\ Schackert}
\affiliation{Physikalisches Institut, Karlsruhe Institute of Technology (KIT), Germany}

\author{T.\ Miyamachi}
\affiliation{Physikalisches Institut, Karlsruhe Institute of Technology (KIT), Germany}
\affiliation{\blue{Institute of Solid State Physics, University of Tokyo, Japan}}

\author{F.\ Schramm}
\affiliation{Institute of Nanotechnology, Karlsruhe Institute of Technology (KIT), Germany}

\author{O.\ Fuhr}
\affiliation{Institute of Nanotechnology, Karlsruhe Institute of Technology (KIT), Germany}

\author{R.\ Chandrasekar}
\affiliation{School of Chemistry, University of Hyderabad, India}

\author{M.\ Alouani}
\affiliation{Institut de Physique et Chimie des Mat{\'e}riaux de Strasbourg (IPCMS), Strasbourg, France}

\author{M.\ Ruben}
\affiliation{Institute of Nanotechnology, Karlsruhe Institute of Technology (KIT), Germany}
\affiliation{Institut de Physique et Chimie des Mat{\'e}riaux de Strasbourg (IPCMS), Strasbourg, France}

\author{W.\ Wulfhekel}
\affiliation{Physikalisches Institut, Karlsruhe Institute of Technology (KIT), Germany}
\affiliation{Institute of Nanotechnology, Karlsruhe Institute of Technology (KIT), Germany}

\author{F.\ Evers}
\affiliation{Institut f{\"u}r Theorie der Kondensierten Materie, Karlsruhe Institute of Technology, Germany}
\affiliation{Institute of Nanotechnology, Karlsruhe Institute of Technology (KIT), Germany}

\date{\today}

\begin{abstract}
We report a combined experimental and theoretical study of the 
Kondo effect in a series of binuclear metal-organic complexes of the form 
[(Me(hfacac)$_2$)$_2$(bpym)]$^0$, with Me = Nickel (II), Manganese(II), Zinc (II); 
hfacac = hexafluoroacetylacetonate, and bpym = bipyrimidine,
adsorbed on Cu(100) surface. 
While Kondo-features did not appear in the scanning tunneling spectroscopy 
spectra of {non-magnetic} Zn$_2$, a zero bias resonance was resolved in magnetic 
Mn$_2$ and Ni$_2$ complexes. The case of Ni$_2$ is particularly interesting 
as the experiments indicate two adsorption geometries with very different properties. 
For Ni$_2$-complexes we have employed  
density functional theory to further elucidate the situation. 
Our simulations show that one geometry with relatively large Kondo 
temperatures $T_\mathrm{K} \sim 10$~K can be attributed to distorted Ni$_2$ complexes, 
which are chemically bound to the surface via the \-bi\-pyri\-mi\-dine unit. 
The second geometry, we assign to molecular fragmentation: we suggest that the original 
binuclear molecule decomposes into two pieces, including 
Ni(hexa\-fluoro\-acetyl\-aceto\-nate)$_2$, when brought into contact with 
the Cu-substrate. For both geometries our calculations support a picture of the 
($S$=1)-type Kondo effect emerging due to open 3$d$ shells of the 
individual Ni$^{2+}$ ions.
\end{abstract}

\maketitle

\section{Introduction}

Molecular electronics holds the vision that functional 
electronic devices, like memory elements, rectifiers and transistors, 
may be realized by designing suitable molecular complexes.
Over the past decade, two- and three-terminal 
molecular junctions with current-voltage characteristics resembling
diod-,\cite{Reichert2002,diods} transistor-\cite{Park2002,Kubatkin2003,Osorio2007} 
or memory-like\cite{Riel2006,Meded2009} behavior have been demonstrated. 
The function of molecular devices can be extended using
the spin degree of freedom initiating the field of molecular spintronics.
An important step in this direction was made recently, when a 
giant magnetoresistance effect has been demonstrated for single molecules 
deposited on a ferromagnetic surface \cite{Schmaus2011,BagretsNanoLett2012}.

In order to achieve external control of electron spins, 
"spin-transition" complexes have been proposed, which can be addressed by 
temperature, pressure, light and presumably 
electron charging \cite{spin-trans-switching,Meded2011}.
In these systems a \textit{\mbox{single} spin} is the main protagonist. When it is 
brought in contact with a substrate, the Kondo-effect sets in. 
It has been observed in systems with increasing complexity ranging from 
3$d$ adatoms on metallic surfaces,\cite{Knorr2002,Berndt2007} 
through molecules with extended 
$\pi$-orbitals,\cite{Gao2007,Fu2007,Pascal2008,Hla2010} 
to carbon nanotubes \cite{Kondotubes1,Kondotubes2}.

The motivation of our work is the question what happens if the 
molecule approaching the surface has \textit{more} 
than a single active spin, say two exchange coupled spins that 
anticipates the case of single molecule magnets (SMMs) \cite{SMMs}.
Interesting new aspects can enter already on the level of 
binuclear magnets \cite{Ruben2013}
and arise from the competition of 
substrate-effects, such as Kondo-screening, 
and the inter-molecular exchange coupling. 

\begin{figure}[t]
\begin{center}
\includegraphics[width=1.0\columnwidth]{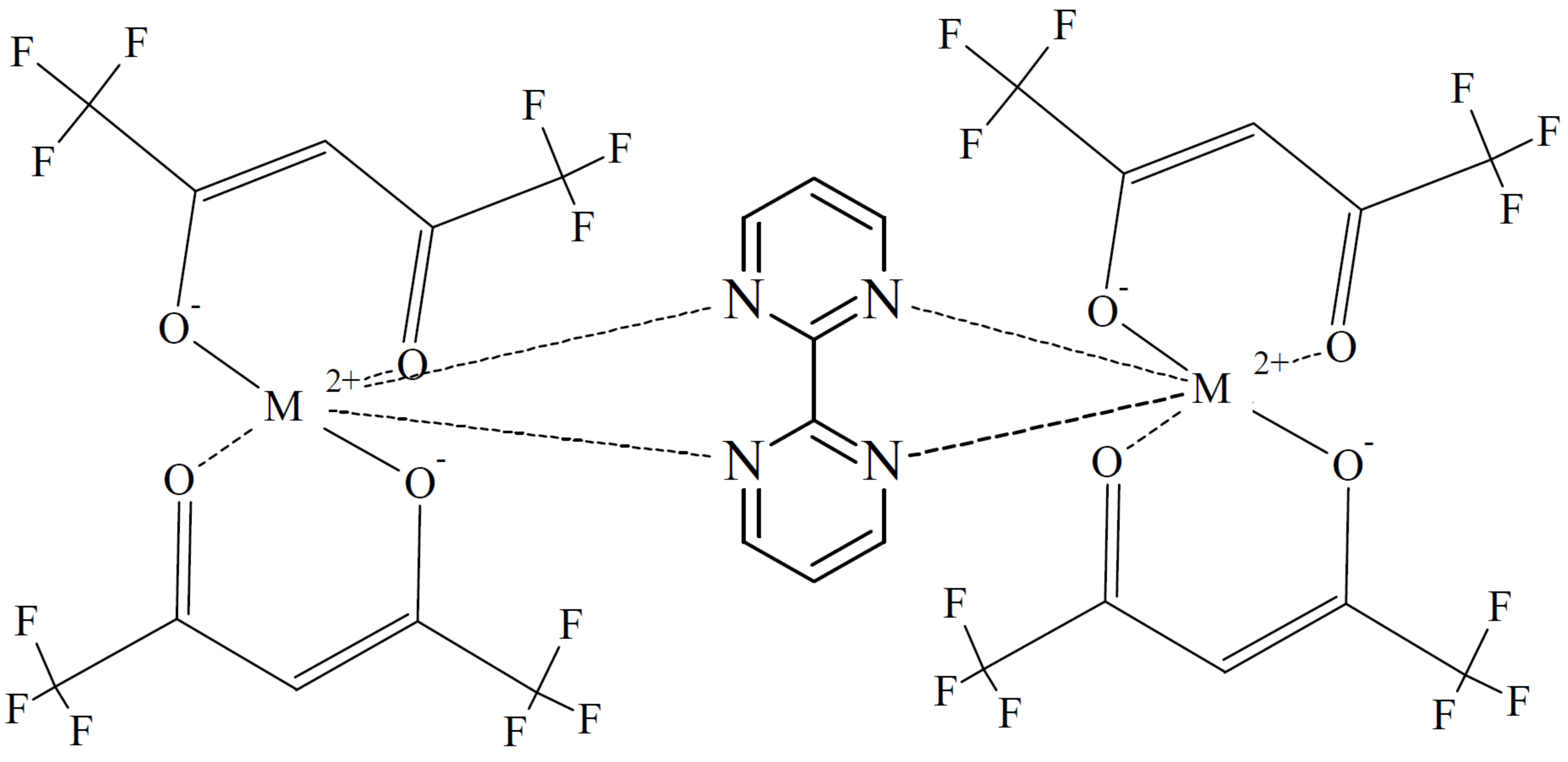}
\caption{Structure of the (Me(hfacac)$_2$)$_2$(bpym) complex. 
Each metal ion Me$^{2+}$ is linked with two hexa\-fluoro\-acetyl\-aceto\-nate
(hfacac) ligands by Me-O bonds. By forming N-Me bonds the
aromatic 2,2'-bi\-pyri\-mi\-dine (bpym) ligand coordinates as bidentate chelate 
with two of the Me(hfacac)$_2$ components.}
\label{Exp_Fig1}
\end{center}
\end{figure}

\begin{figure}[t]
\begin{center}
\includegraphics[width=.95\columnwidth]{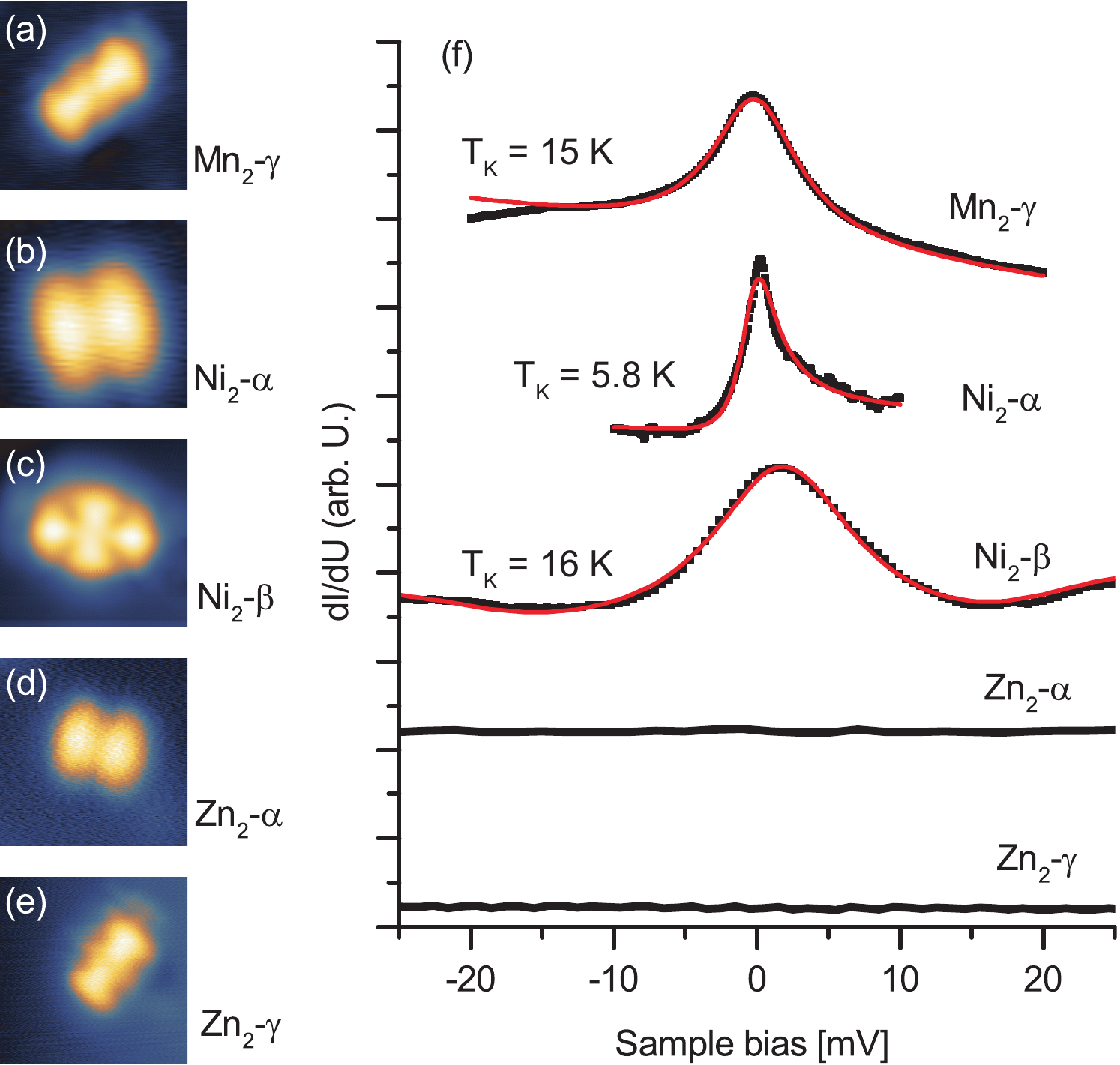}
\caption{
Topography of (a) Mn$_2$-$\gamma$, (b) Ni$_2$-$\alpha$, (c) Ni$_2$-$\beta$, 
(d) Zn$_2$-$\alpha$, and (e) Zn$_2$-$\gamma$ (images a-e are squares of side 
length of 3.7, 2.8, 2.5,  2,9, and 3.3~nm, respectively. Feedback conditions 
are 100~mV and 1~nA, 200~mV and 200~pA, $-10$~mV and 200~pA, 15~mV and 1~nA, 
100~mV and 1~nA, respectively).  
The orientation of $\beta$ and $\gamma$ configuration follows [001] and [011] 
direction of the substrate, respectively, while the orientation of $\alpha$ 
configuration seems to be random. The differential conductance $dI/dU$ 
near the Fermi-level is shown in (f). 
The black dots are experimental data which were fitted with Fano-functions 
(red lines). The STM images and STS were measured at 4.2 K. The spectra are 
normalized and a linear background was considered to obtain reasonable 
fitting result.
}
\label{Exp_Fig2}
\end{center}
\end{figure}

Motivated by such a question, 
we have synthesized a series of binuclear metal-organic complexes of the form 
[(Me(hfacac)$_2$)$_2$(bpym)]$^0$,
referred to later on as "Me$_2$" 
(see Fig.~\ref{Exp_Fig1}). 
These complexes have been deposited on a clean Cu(001) surface and studied with 
low-temperature scanning tunneling microscopy (STM). While no feature of 
the Kondo effect was found in the scanning tunneling spectroscopy (STS) 
spectrum of Zn$_2$ with closed shell 3$d$ ions, an adsorption-site dependent 
zero-bias (Kondo) resonance was clearly resolved in the case of Mn$_2$ 
and Ni$_2$ molecules (see Fig.~\ref{Exp_Fig2}).

The Ni$_2$-case is particularly interesting, as the adsorbed molecule 
appears in two variants that differ in STM-images and STS-characteristics. 
To rationalize these observations, we have performed calculations based on density functional 
theory (DFT). We thus identify possible adsorption geometries.
Our simulations show that one observed geometry (with relatively 
large Kondo temperature $T_\mathrm{K}$) can be attributed to the 
original Ni$_2$ complex with a distorted geometry bound to Cu(001) 
via the bipyrimidine (bpym) unit. The other is likely to  arise from
molecular fragmentation. We  propose that upon approaching the substrate the molecule 
breaks and forms two Ni(hfacac)$_2$ moieties, which are then seen in the experiment. 

\section{Synthesis and characterization of binuclear complexes}

{\bf Molecular geometries: } The molecular structure of the neutral binuclear metal complexes 
[(Me\-(hfacac)$_2$)$_2$\-(bpym)]$^0$, with Me = Nickel(II), Man\-ga\-nese(II), Zinc(II); hfacac = 
hexa\-flu\-o\-ro\-acetyl\-ace\-to\-nate, and bpym = bi\-py\-ri\-mi\-dine, 
studied in this work is presented 
in Fig.~\ref{Exp_Fig1}. The structural details of the metal complexes, in particular the coordination 
environment of the metal ions, are sensitive to the kind of 3$d$ metal ion 
involved (for details, see Supplementary Information,\cite{suppinfo} Sec.~IV). 
According to single crystal X-ray dif\-frac\-to\-met\-ry each Mn$^{2+}$ ion is situated 
in a distorted trigonal prismatic N$_2$O$_4$ coordination sphere leading to a Mn-Mn distance 
of 6.2~\AA\ in Mn$_2$ dimer. By contrast, the Ni$_2$ complex exhibits a distorted octahedral 
coordination environment of each 
Ni$^{2+}$-ion \cite{Barquin1999} with a Ni-Ni distance of only 5.6~\AA. 

{\bf Magnetism:} The magnetic behavior of polycrystalline samples of the
Mn$_2$ and Ni$_2$ complexes was determined between 2~K and 300~K
(Supplementary Information,\cite{suppinfo} Sec.~IV). At room
temperature Mn$_2$ has a $\chi_MT$ value of 9.03~cm$^3${$\cdot$}K{$\cdot$}mol$^{-1}$
(here $\chi_M$ is the molar magnetic susceptibility and $T$ is the temperature),
corresponding to two uncoupled high-spin Mn(II)-ions with a spin value of $S{=}5/2$ each, 
while Ni$_2$ shows a $\chi_MT$ product of 2.23~cm$^3${$\cdot$}K{$\cdot$}mol$^{-1}$,
corresponding to two uncoupled Ni(II)-ions with $S{=}1$. Between 300~K and 75~K both the 
Mn$_2$ and Ni$_2$ complexes show paramagnetic behavior, while below 75~K weak
antiferromagnetic (AF) behavior sets in. The AF exchange interaction
between the two divalent 3$d$ ions through a bpym bridging ligand is
reported as $J^\mathrm{AF}_\mathrm{ex} = 1.6$~meV \cite{Brewer1985}
and $J^\mathrm{AF}_\mathrm{ex}= 2.06$~meV \cite{Barquin1999} 
for Ni$_2$, which is about 10 times larger than the exchange interaction
observed in Mn$_2$ with $J^\mathrm{AF}_\mathrm{ex} = 0.13$~meV.

{\bf DFT-calculations for the Ni$_2$ complex:} 
The spin state of each of the two Ni(hfacac)$_2$ units was determined as 
$S=1$ assuming an [Ar]$3d^{8}$ electronic configuration for Ni$^{2+}$ ion. 
Specifically, the atomic structure of the non-adsorbed (gas-phase) 
Ni$_2$ complex was refined starting form the X-ray diffraction structure with
Ni$^{2+}$ ions in distorted octahedral environment 
(see Supplementary Information\cite{suppinfo}, Sec.~I, for computational details).
The results of the "constrained" DFT calculations are summarized in the 
first row of Table~I. In particular, our calculations 
predict a "singlet" ground state with the two spins, $S{=}1$, 
coupled antiferromagnetically. 
For the gas-phase complex, very low excitation energy 
($\simeq 15$~meV) is observed with an excited state exhibiting 
ferromagnetic coupling. Our results
are consistent with experimental findings: weak antiferromagnetic 
exchange interaction between the two metal ions of the
[(Me(hfacac)$_2$)$_2$(bpym)]$^0$ complexes facilitated by the bpym 
ligand has been reported before \cite{Barquin1999}. 

\section{STM experiments}

The STM experiments were performed using a home-built, low-noise 
STM operating between 0.7--4.2~K in ultra-high vacuum (UHV) ($p<10^{-9}$~mbar) \cite{Zhang2011}. 
Clean and atomically flat Cu(100) substrates were prepared \textit{in situ} followed by deposition of molecules 
by sublimation at 80--100~$^{\circ}$C. Three adsorption configurations, $\alpha$ (Ni$_2$ and Zn$_2$), 
$\beta$ (only Ni$_2$) and $\gamma$ (Mn$_2$ and Ni$_2$), of the molecules on Cu(100) 
were found (see Fig.~\ref{Exp_Fig2}a-e). Scanning tunneling spectroscopy (STS) 
measurements were performed on these different 
configurations. Kondo-like peaks were clearly resolved in 
the $dI/dU$ curves of Mn$_2$ and Ni$_2$ (in both configurations) near the Fermi-level  
while no remarkable feature was found in the $dI/dU$ curve of the non-magnetic 
Zn$_2$ complexes (see spectra displayed in Fig.~\ref{Exp_Fig2}f).

The Kondo-effect arises when the magnetic moment of an impurity is screened by 
surrounding electrons of a nonmagnetic substrate \cite{Anderson1961,Kouwenhoven2001}. 
As the simplest manifestation 
of the interaction between the localized spin and delocalized electrons, the 
shape of the zero bias anomaly in STS caused by the Kondo-effect can be 
described by a Fano resonance \cite{Fano1961,Ujsagny2000,Merino2004}:

\begin{equation}
\frac{dI}{dU}(U) \propto \frac{(\varepsilon+q)^2}{1 + \varepsilon^2},
\label{eq1}
\end{equation}
where
\begin{equation}
\varepsilon = \frac{eU - \varepsilon_0}{\Gamma},
\label{eq2}
\end{equation}
and $\varepsilon_0$ is the energy shift of the resonance from the Fermi-level, 
$\Gamma$ is the width of the resonance. The Fano parameter $q$ characterizes the 
interference of tunneling between the tip and the magnetic impurity and 
tunneling between the tip and the sample \cite{Plihal2001}. Considering the temperature 
dependence, the Kondo-resonance can be approximated by a Lorentzian resonance \cite{Nagaoka2002}. 
Thus the energy width $2\Gamma(T)$ (full width at half maximum) 
of a Kondo-resonance can be expressed as: 
\begin{equation}
2\Gamma(T) = 2\sqrt{(\pi k_\mathrm{B} T)^2 + 2(k_B T_\mathrm{K})^2},
\label{eq3}
\end{equation}
with $k_\mathrm{B}$ being the Boltzmann constant, $T$ being the environment temperature, 
and $T_\mathrm{K}$ being the Kondo-temperature. By fitting the experimental STS 
with equations (\ref{eq1})--(\ref{eq3}), the Kondo-temperature $T_\mathrm{K}$ is extracted. 

\begin{figure}[b]
\begin{center}
\includegraphics[width=0.95\columnwidth]{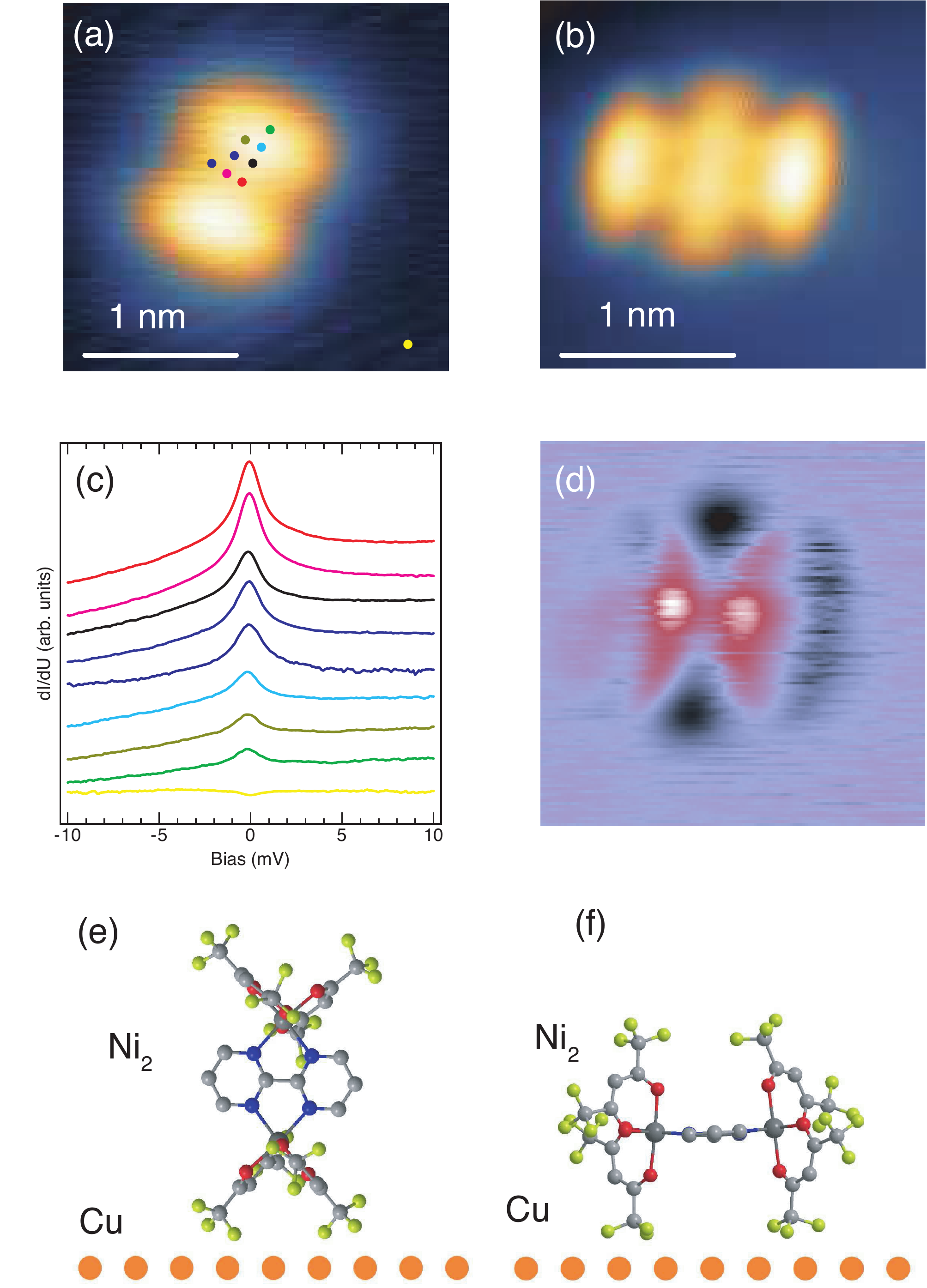}
\caption{
Topography of Ni$_2$-$\alpha$ (a) and Ni$_2$-$\beta$ (b). 
STS on different position of the Ni$_2$-$\alpha$ 
is shown in (c) ($T=1$~K, feedback conditions: $U=10$~mV, $I=20$~nA, 
$200~\mu$V modulation). STS $(dI/dU)$ map of Ni$_2$-$\beta$ is shown in (d). 
White areas indicate large differential conduction, black low. ($T=5$~K, 
feedback conditions $U=10$~mV, $I=10$~nA). 
An intuitive assumption for the adsorption 
configurations of Ni$_2$-$\alpha$ and Ni$_2$-$\beta$ (front view) 
on the Cu(100) substrate is shown in (e) and (f), respectively.
}
\label{Exp_Fig3}
\end{center}
\end{figure}

The Zn$^{2+}$ ions in the complex are expected to have a full 3$d$ sub-shell, 
so a magnetic moment is absent. This is in agreement with the SQUID measurements 
of the crystals of this complex. Thus, a Kondo-effect could appear only in adsorbed 
molecules if a charge transfer between the molecule and the substrate leads to the 
acquisition of a magnetic moment on the molecule. Since our measurement do not indicate 
a Kondo-resonance (compare Fig.~\ref{Exp_Fig2}f), we conclude that the interaction 
with the substrate is too weak for such a charge transfer. 
Contrary to Zn$^{2+}$, in  Mn$_2$ and Ni$_2$  
the central Mn$^{2+}$/Ni$^{2+}$ ions exhibit a partially 
filled 3$d$ shell and therefore carry a finite 
magnetic moment. They are at the origin of 
the Kondo-effect that we observe in our measurements, Fig.~\ref{Exp_Fig2}. 
By fitting to the Fano-shaped resonance, a Kondo-temperature of 15~K 
for Mn$_2$ is determined.

With Ni$_2$ the situation is more complex since 
two different adsorption geometries (referred to as $\alpha$ and $\beta$) 
are observed, see Fig.~\ref{Exp_Fig2}b,c. 
We determine two Kondo-temperatures, 5.8~K and 16~K for Ni$_2$-$\alpha$ 
and Ni$_2$-$\beta$, respectively. 
An even clearer difference between these adsorption geometries exhibits itself in a 
site-dependent STS measurement that we perform on Ni$_2$-$\alpha$ and Ni$_2$-$\beta$, 
see Fig.~\ref{Exp_Fig3}. The measurement shows that the Kondo-resonance has a single maximal 
amplitude at the center of the Ni$_2$-$\alpha$ complex (cf. Fig.~\ref{Exp_Fig3}a,c), 
while two spots with maximal amplitude 
of $dI/dU$ signal separated by a distance of 4~\AA\ are clearly resolved in the 
STS map of Ni$_2$-$\beta$ (cf. Fig.~\ref{Exp_Fig3}b,d), which match 
with the two Ni$^{2+}$ ions and their expected distance in the molecule.

At first sight, these results suggest that 
adsorption configurations of Ni$_2$-complexes may look 
like shown in Fig.~\ref{Exp_Fig3}(e,f). 
Namely, Ni$_2$-$\beta$ could correspond to the molecular complex, which `lies' on 
the surface with both of its hfacac ligands, thus exposing the two Ni ions seperately to the STM tip. 
The $\alpha$-configuration could correspond to the 
complex, which `stands' on the surface with one hfacac ligand and the other hfacac ligand
is seen in its topographic image. This would expose the two Ni ions above each other such 
that a single Kondo-resonance is observed in STS. 
As we show later on, the latter assignment is inconsistent with 
the theoretical considerations. Below, we propose an alternative scenario. 
\bigskip

\section{Theory}

To elucidate microscopic details of the Kondo-effect observed
experimentally, we have performed elaborated DFT calculations. 
Our main objective is to understand the dependence of 
the Kondo-resonance of the Ni$_2$ complex on the adsorption site. 
Additional questions that we address will concern the nature of the 
molecular orbitals involved in the interaction with 
conduction electrons, and how delocalized electrons compete
for screening of the initially antiferromagnetically coupled spins.

\subsection{Simple adsorption geometries}

We analyze the adsorption geometries schematically 
illustrated in Fig.~\ref{Exp_Fig3}(e,f). They exhibit 
gas-phase Ni$_2$ complexes placed on Cu(001) 
surface (see Supplementary Information,\cite{suppinfo} Sec.~IIA, 
for computational details). 
Our simulation results indicate fluorine-copper distances above $\sim 3$~\AA. 
This distance implies that a weak van-der-Waals (vdW) force dominates binding 
to the surface. There is only a weak hybridization between molecular and 
substrate states, which translates into narrow molecular resonances, 
$\Gamma\simeq 10$~meV, as seen from the spectral function $A(E)$ 
projected on the Ni(II) ion (see Supplementary Information,\cite{suppinfo} Suppl.\ Fig.~2).
Giving typical parameters of the Anderson 
model read from $A(E)$, namely, single occupied resonance level width 
$\Gamma \simeq 10^{-2}$eV, on-site Coulomb repulsion energy
$U \sim 2$eV, and position of the resonance level  $\varepsilon_d \sim U/2 \sim 1$eV relative to 
the Fermi-energy, we can estimate the Kondo-temperature as~\cite{Hewson,Hewson1}
$k_B T_\mathrm{K} \sim U \sqrt{\frac{\Gamma}{4U}}\, e^{-\pi U/4\Gamma} 
\sim 0.1\, e^{-10^2\pi/2}~\mathrm{eV} \sim 10^{-69}$~eV. This result is contradicting
the experimental finding of $k_B T_\text{K}\sim 10$~meV, so that we exclude 
adsorption geometries shown in Figs.~\ref{Exp_Fig3}(e,f). 
Alternatives should allow for a stronger binding with a 
significant amount of hybridization in order to achieve larger 
$\Gamma$-values. 

\begin{figure}[b]
\begin{center}
\includegraphics[width=1.0\linewidth]{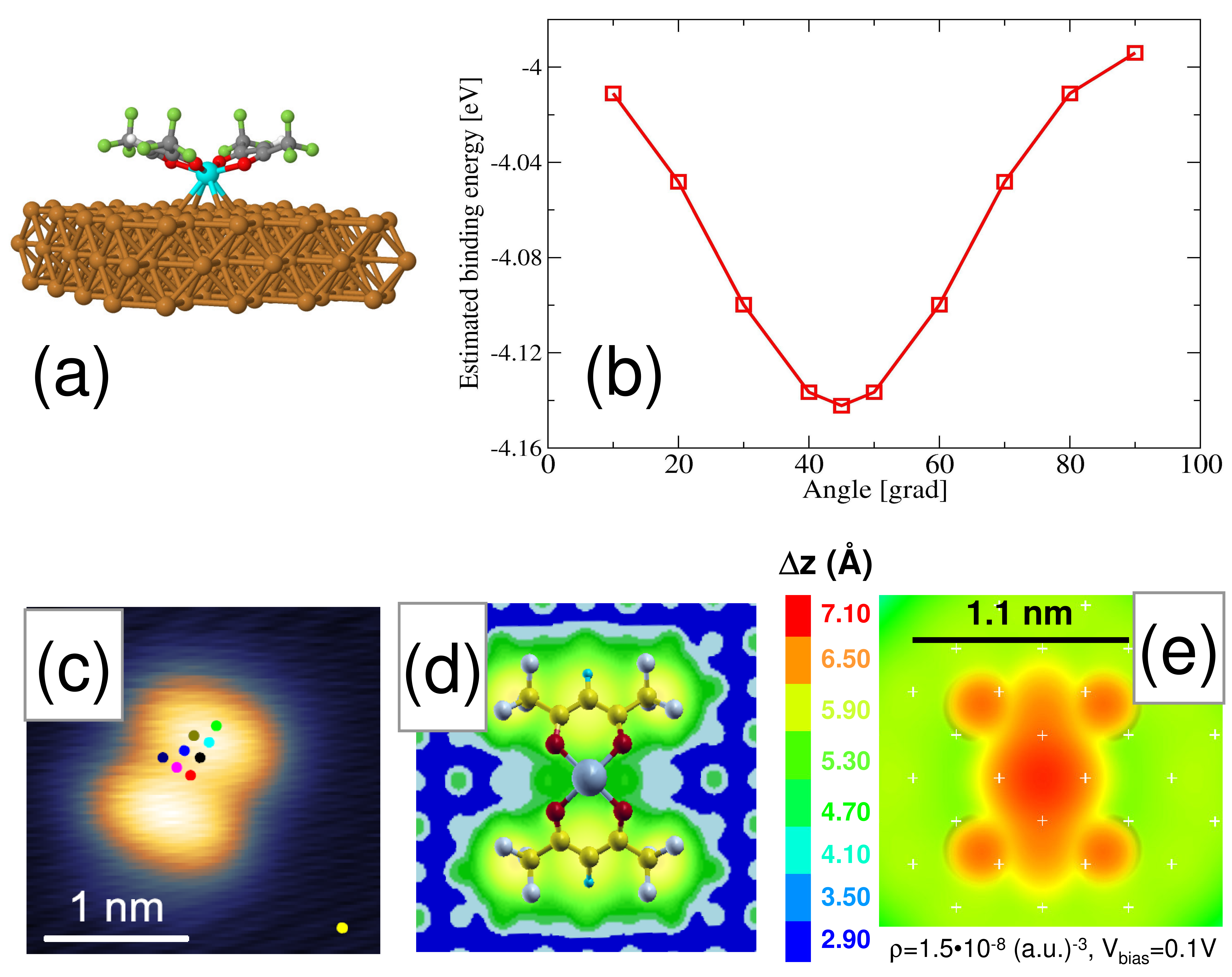}
\caption{
(a) Ni(hfacac)$_2$ moiety bound to a Cu(001) surface via a Ni atom 
placed at the {\it hollow} site. (b) Estimated dependence of the
binding energy of Ni(hfacac)$_2$ to Cu (001) surface on the 
angle that fixes the orientation of the moiety's mirror planes 
\textit{vs} the [001] direction of the fcc (001) surface.
(c) Experimentally recorded Ni$_2$-$\alpha$ STM image (reproduced
from Fig.~\ref{Exp_Fig2}).
(d) Simulated image of Ni(hfacac)$_2$, computed employing VASP package \cite{VASP}:
different colors refer to isosurfaces of the space-resolved 
(local) density of states (DoS) at the Fermi-level. 
(e) Simulated image of Ni(hfacac)$_2$, computed employing 
{\small AITRANSS} package\cite{aitranss1,aitranss2}: selected 
three-dimensional isosurface of the space-resolved 
DoS (above the molecule) integrated over the energy window 0.1 eV 
around the Fermi-level is shown, where color coding refers to the 
distance $\Delta z$ to the Cu surface. In consistent with experimental
STM image, simulated images obey two mirror planes and 
outermost contours with a butterfly shape.}
\label{Fig_MoietyGeo}
\end{center}
\end{figure}

\begin{table*}[t]
\begin{center}
\caption{%
Relative energies of the low-energy spin configurations realized in Ni$_2$ complexes, 
which were estimated based on the ground state constrained DFT calculations
(see Suppl.\ Information\cite{suppinfo} for computational details).
Considered molecules are: 
(i) gas-phase relaxed Ni$_2$ complex, with Ni$^{2+}$ ions found in 
distorted octahedral environment (Suppl.~Fig.~1a);
(ii) free but distorted Ni$_2$ complex, pre-relaxed in the presence of Cu surface, 
where local $C_{2v}$ symmetry has been kept (Suppl.~Fig.~4a); 
(iii) free but distorted Ni$_2$ complex, relaxed in the presence of Cu surface 
where symmetry constrains have been released (Suppl.~Fig.~4b).
Spin configurations labeled as F and AF stand for ferromagnetic (F) and antiferromagnetic (AF) coupling, 
respectively, between $S=1$ spins localized on Ni$^{2+}$ ions.
Last column refers to magnetically excited zero-spin (closed-shell) state
at one of Ni$^{2+}$ ions.
}
\begin{ruledtabular}
\begin{tabular}{cccccccc}
{} & {DFT implementation} & & 
AF:~$\uparrow\uparrow~\downarrow\downarrow$ & & F:~$\uparrow\uparrow~\uparrow\uparrow$ & & $\uparrow\uparrow~\uparrow\downarrow$ \\
\hline
(i) relaxed Ni$_2$ & {\small{FHI}}-aims\cite{fhi-aims}   & &  0.000~eV   & &   15.8~meV  & &   0.500~eV  \\ 
      complex      & {\small{TURBOMOLE}}\cite{turbomole} & &  0.000~eV   & &   14.9~meV  & &   0.522~eV  \\ 
\hline
(ii) distorted Ni$_2$ complex  & {\small{FHI}}-aims   & &  0.000~eV    & &   6.9~meV  & &   0.396~eV  \\ 
with $C_{2v}$ symmetry       & {\small{TURBOMOLE}}  & &  0.000~eV    & &   6.1~meV  & &   0.390~eV  \\ 
\hline
(iii) distorted      & {\small{FHI}}-aims   & &  0.000~eV    & &   2.5~meV  & &   0.456~eV  \\ 
Ni$_2$ complex       & {\small{TURBOMOLE}}  & &  0.000~eV    & &   1.9~meV  & &   0.361~eV  \\ 
\end{tabular}
\end{ruledtabular}
\end{center}
\end{table*}

\subsection{$\alpha$-configuration: Ni(hfacac)$_2$ fragments on Cu(001)}

In order to enforce a much larger coupling of the molecular complex 
to the substrate we first consider the extreme case of molecular fragmentation,
see Fig.~\ref{Fig_MoietyGeo}a. Here we include the possibility that 
coordination bonds between Ni$^{2+}$ ion and nitrogen atoms are broken, 
and a Ni(hfacac)$_2$ moiety, which is chemically bound to a Cu surface,
is observed in the experiment. 
In that situation the coupling of the spin to the Cu surface 
is comparable to the case of single Ni adatom (no ligands attached), 
and drawing from related earlier experimental 
experience \cite{Knorr2002,Berndt2007,Jamneala} 
one might suspect Kondo-temperatures of the order of tenth of meV roughly consistent with 
the present measurements.  
In order to show that the fragmentation scenario is 
consistent with the experimental findings, we observe following facts: 

(i) \textit{Simulated STM-images.} Our simulations of fragmented molecules 
(Fig.~\ref{Fig_MoietyGeo}d,e) yield STM-images reproducing the most important 
characteristics of the experimental ones for Ni$_2$-$\alpha$: the outermost contours 
have a butterfly shape, two mirror planes exist, the size of experimental 
and computational images are consistent. 
It is encouraging to see that also non-trivial details are (partially) reproduced. 
Namely, theory predicts a non-zero optimal angle, 45$^\circ$ 
(Fig.~\ref{Fig_MoietyGeo}b) that fixes orientation of the fragment's mirror planes 
\textit{vs} the fcc [100] direction of the (001) surface plane
(see also Supplementary Information,\cite{suppinfo} Suppl.\ Fig.~3). 
A non-zero angle, $\approx$~20$^\circ$, is also 
observed in the representative experimental images of Ni$_2$-$\alpha$
(see Suppl.\ Fig.~5).

(ii) \textit{Spatial dependency of Kondo-amplitude.}
The fragment's geometry 
is such that the associated Kondo-resonance would have 
maximum amplitude with the STM-tip located in the center of 
the image (at Ni atom). That is in consistent with the
structure of the spatially resolved Kondo-resonance
of Ni$_2$-$\alpha$ observed experimentally (cf.\ Fig.~\ref{Exp_Fig2}a,c).

(iii) \textit{The effect of temperature on the adsorption.}
Experimental STM images characterizing adsorption 
of Ni$_2$ complexes on Cu(001) surface (Suppl.\ Fig.~5) 
suggest that molecular fragmentation
at surface is likely triggered by the temperature.  
Namely, experimental data (Suppl.\ Fig.~5) reveal that 
when Ni$_2$ complexes are evaporated
on the substrates at room temperature, two species -- Ni$_2$-$\alpha$ and 
Ni$_2$-$\beta$ -- are found, while Ni$_2$-$\alpha$, which we
attribute to molecular fragments, is not found after deposition onto pre-cooled 
substrates (77K).

\begin{figure}[t]
\begin{center}
\includegraphics[width=0.9\linewidth]{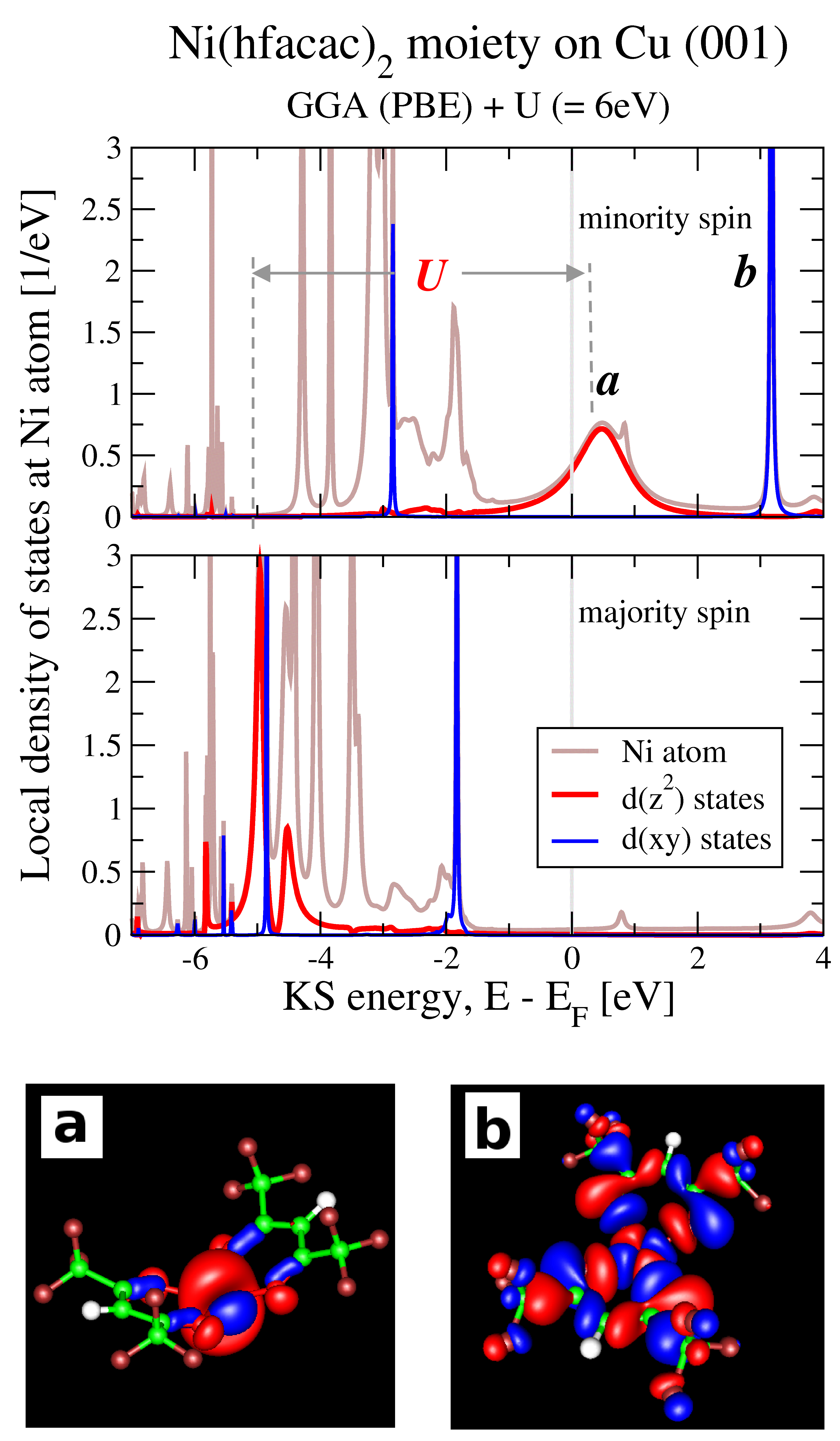}
\caption{%
Upper plot: spin dependent local density of states projected on Ni atom 
of the Ni(hfacac)$_2$ on Cu(001). Red and blue lines highlight contributions 
to the LDOS associated with Ni $d_{z^2}$ and $d_{xy}$ orbitals. Lower plots: 
corresponding Kohn-Sham wave functions of the free standing molecule. }
\label{Fig_MoietySpectrum}
\end{center}
\end{figure}

Since the fragmentation scenario is consistent with the experimental 
phenomenology, we perform further \textit{ab initio} calculations with 
the goal to better understand the molecular magnetism and eventually 
estimate the Kondo-temperature. 

We consider the atomic configuration of Ni atom as [Ar]$4s^23d^8$.
In the simplified picture of a free Ni(hfacac)$_2$-fragment, 
the 4$s$-states hybridize so strongly with the ligands, that the 
$s$-electrons are effectively transferred to ligand orbitals. Therefore, 
the metal ion takes the Ni$^{2+}$-configuration and exhibits two unpaired spins. 
We performed a DFT-study within the generalized gradient 
approximation (GGA, PBE exchange-correlation functional \cite{pbe})
of the molecule in gas-phase (for details, see Supplementary Information,\cite{suppinfo} Sec.I).
Our results confirm above picture: we 
find a spin-polarized ground state with a 
magnetic moment of $2\mu_B$. The magnetization is largely due to 
two orbitals (\textit{a} and \textit{b} on Fig.~\ref{Fig_MoietySpectrum}) 
that are populated with up-spin electrons, only, and 
that contribute substantial weight to both Ni $d_{z^2}$ and $d_{xy}$ atomic states. 

One may ask whether the fragment keeps its magnetic moment when 
adsorbed on the substrate. To answer this question, 
we performed another spin-DFT study, whose details 
are presented in the Supplementary Information\cite{suppinfo} (Sec.II and Suppl.\ Fig.~3). 
In essence, the substrate further breaks the residual 
degeneracy of the $d$-orbitals splitting $d_{z^2}$ and $d_{xy}$ by 1.4~eV. 
As a result, $d_{z^2}$ is nearly full (0.4~eV below $E_\text{F}$), 
while $d_{xy}$ is nearly empty (1~eV above $E_\text{F}$ with resonance 
$\Gamma_b\approx 10^{-2}$eV). The magnetization drops after 
adsorption by about a factor of two, down to 1.2 $\mu_B$. 
The net charge(electron)-transfer to the molecule in the adsorption process is 0.4$e$. 

The GGA-type calculation leaves us with a situation where a localized 
$3d_{z^2}$-type orbital is only partially occupied. 
In DFT with conventional local exchange-correlation (XC) functionals
(LDA, GGA) this is typically an indication of correlation physics, such as the 
Coulomb-blockade. The Coulomb-blockade is not described by conventional 
DFT-functionals, but it can be captured on the level 
of GGA+U. \cite{Anisimov1997,KuemmelKronik2008}
Therefore, following Ref.~\onlinecite{Dudarev}, 
we performed a GGA+U calculation 
placing a repulsive on-site term with (relatively large)
strength $U =6$~eV on the metal site. 
(Details of our implementation are given in Ref.~\onlinecite{aitranss3}.)
The interaction shifts the spin-down resonance \textit{a} 
from 0.4 eV below to 0.5eV above $E_F$ (Fig.~\ref{Fig_MoietySpectrum}). 
Hence, the magnetic moment increases up to 1.62~$\mu_B$ thus suggesting 
the picture of the ($S$=1) Kondo-effect. 

So far our \textit{ab initio} study has ignored the spatial structure 
of the two molecular orbitals involved, 
orbital $a$ with substantial contribution from Ni $d_{z^2}$ atomic state, 
and orbital $b$ with substantial contribution from  Ni $d_{xy}$ atomic state.
These molecular orbitals are depicted in Fig.~\ref{Fig_MoietySpectrum}. 
As can be seen, the \textit{a}-orbital (first quantum dot, $\Gamma_a$)
is directed towards the surface. Hence, it hybridizes 
with the substrate much stronger than the \textit{b} orbital (second dot $\Gamma_b$), 
i.e.\ $\Gamma_b \ll \Gamma_a$. Each level has a single occupancy and the 
electrons populating them are coupled ferromagnetically.
Since the exchange interaction ($\sim 0.5$~eV, see Table~1) is much larger compared 
to the expected Kondo energy scale $\sim 10$~K, both spins form a triplet, $S=1$. 
Reading parameters from the spectral function (Fig.~\ref{Fig_MoietySpectrum}), 
\[
\Gamma_a \simeq 0.8~\mathrm{eV} \gg \Gamma_b;\ 
\varepsilon_d \simeq 0.5~\mathrm{eV}; \ 
U \simeq 5.5~\mathrm{eV} \gg \varepsilon_d,
\]
and using a formula \cite{Hewson,Hewson1} for the Kondo-temperature, 
\begin{equation}
k_B T_\mathrm{K} \simeq 0.41\, U \sqrt{\frac{\Gamma^*}{4U}}\, e^{-\pi \varepsilon_d/\Gamma^*},
\label{eqTK}
\end{equation}
where $U \gg \varepsilon_d$ and $\Gamma^* = \Gamma_a/2$ for the case of double-dot system,  
we obtain $k_B T_\mathrm{K} \simeq 5.84 \times 10^{-3}~\mathrm{eV}$, 
i.e.\, a Kondo-temperature $\simeq 70~\mathrm{K}$ in qualitative agreement with the experiment.

We emphasize that a more precise estimation of the Kondo-temperature is 
hampered by exponential dependence of $T_\mathrm{K}$ on model parameters. 
For example, taking into account that DFT has a tendency to overestimate resonance line-widths,
we may assume a slightly smaller $\Gamma_a \simeq 0.6$~eV. This reduces 
$T_\mathrm{K}$ down to $\simeq 16$~K, which is in good agreement with the 
experimentally observed value. 

\subsection{$\beta$-configuration: distorted Ni$_2$ complex on Cu(001)}

\begin{figure}[b]
\begin{center}
\includegraphics[width=0.95\linewidth]{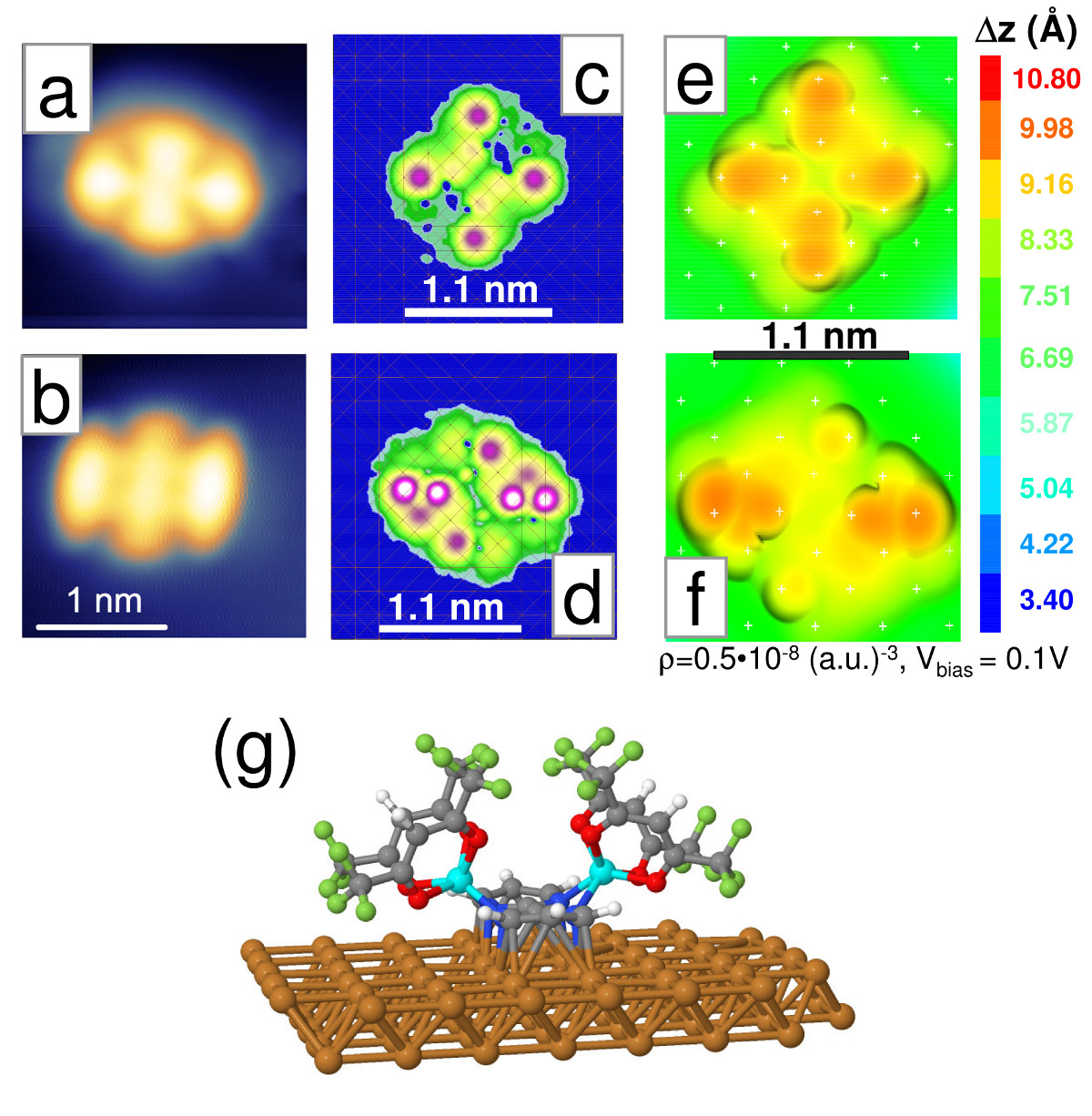}
\caption{Experimentally recorded (a,b) Ni$_2$-$\beta$ and simulated STM images of the
Ni$_2$ complex, with partially weakened chemical bonds, bound to Cu(001) via 
the bpym moiety (see lower plot). Theoretical images are 
computed with VASP \cite{VASP}(c,d) and {\small AITRANSS} \cite{aitranss2}
(e,f). Images (c) and (e) are obtained
assuming symmetry constrains within the DFT relaxation procedure, while images (d) and (f)
correspond to the relaxed structure without constrains 
(see text for further details). Lower plot shows distorted
Ni$_2$ complex with partially weakened chemical bonds is bound to Cu(001) 
via (bpym) moiety.}
\label{Fig_Ni2STM}
\end{center}
\end{figure}

Spatially resolved intensity of the Kondo-resonance, 
measured on top of Ni$_2$-$\beta$ image (Fig.~\ref{Exp_Fig3}d)
suggests that also molecular species with two metal ions are to be found on a Cu surface.
To rationalize this observation, we performed numerical simulations,
to find an intact but strongly distorted molecular conformation (see Fig.~\ref{Fig_Ni2STM}g)
that establishes a chemical bond to the Cu surface ---
an important prerequisite for observable Kondo-temperatures. 
This bond is presumably realized via the delocalized $\pi$-orbitals of the
central bpym unit overlapping with the electron
density extending from the surface. The bond involves contributions
of the vdW forces, which have been accounted for in our simulations,
and presumably has ionic character due to fractional charge transfer to the 
quasi-degenerate \mbox{LUMOs}\cite{lumos}.

We mention that even for the modern \textit{ab initio} methods,
finding the relaxed ground state structure of a large organic molecule (such as Ni$_2$ complex)
on a surface is a non-trivial procedure. Due to the many
atomic degrees of freedom involved, relaxation can end up in different molecular
conformations with energies differing by $\sim$~100 meV, as was 
also the case in our simulations. To be specific in the following 
discussion, we focus on two representative but 
slightly different conformations of "distorted" Ni$_2$ complex 
(for details, see Suppl.\ Fig.~4).

The first conformation (see Suppl.\ Fig.~4a) 
has been obtained within the preliminary DFT 
relaxation procedure: the atomic structure 
of the complex has an (approximate) 
$C_{2v}$ symmetry, in registry with the underlying fcc(001) surface.
For this conformation, the simulated STM image of the complex 
reveals a "cross-like" structure, resembling Ni$_2$-$\beta$
experimental images (see Fig.~\ref{Fig_Ni2STM}a,c,e). 
Further relaxation steps within the simulation
account for an energy gain of about $\sim 0.25$~eV:
the local symmetry of the molecular complex is broken
resulting in the second conformation (see Suppl.\ Fig.~4b). 
Then formerly symmetric simulated "cross-like" STM images are transformed to 
the ones with broken symmetry (cf. Fig.~\ref{Fig_Ni2STM}d,f), 
which were also experimentally observed (cf. Fig.~\ref{Fig_Ni2STM}b).

The structure of the binuclear complex 
suggests that the molecular spins should reside on the Ni(hfacac)$_2$ units, where
each unit could accept two unpaired electrons (referred to as $S$=1) owing
to [Ar]$3d^{8}$ electronic configuration of the Ni$^{2+}$ ion.
Essentially, the two $S$=1 subsystems are magnetically nearly
decoupled, since only a weak indirect ("super-exchange") interaction between them 
could be realized via the $\pi$-orbitals of the bpym unit.
Thus, we anticipate that each subsystem will develop a Kondo-effect, 
independently, as the molecular complex provides two 
parallel conduction paths (channels) for the tunneling electron, one for each spin.

These expectations are fully confirmed by our computational analysis.
Constrained DFT calculations (see data in Table~1) predict 
a ``singlet'' ground state with 
antiferromagnetically (AF) coupled $S$=1 spins. 
For the distorted Ni$_2$ complex, the state with ferromagnetically (F) 
coupled $S$=1 spins is only $\sim 2$~meV above the AF state. 

\begin{figure}[t]
\begin{center}
\includegraphics[width=0.95\linewidth]{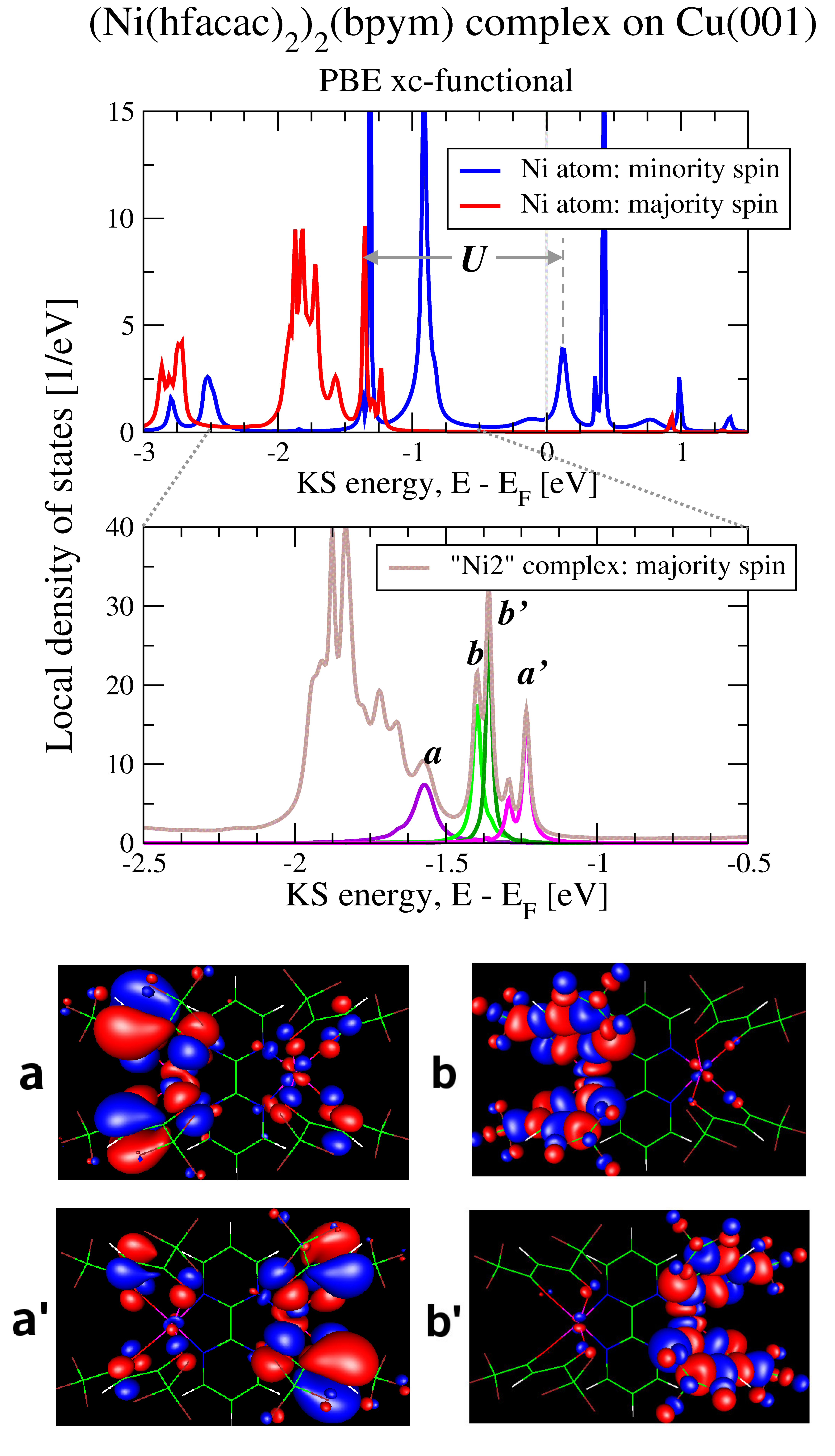}
\caption{\label{Fig_Ni2Spectrum}
Upper plot: spin-dependent local density of states projected on one of Ni atoms of the 
distorted Ni$_2$ complex deposited on Cu(001) [for a geometrical arrangement, 
see Fig.~\ref{Fig_Ni2STM}g]. Middle (zoomed in) plot shows 
majority (up-) spin spectral function below the Fermi-level, 
where contributions are highlighted arising from four orbitals 
(\textit{a}, \textit{b}, \textit{a'} and \textit{b'}), 
each carrying one unpaired spin. Corresponding wave functions are presented below. 
There \textit{a} and  \textit{b} are localized on the "left-hand" side, while
\textit{a'} and \textit{b'} are localized on the "right-hand" side of the molecular complex.
}
\end{center}
\end{figure}

The frontier molecular orbitals (see Fig.~\ref{Fig_Ni2Spectrum}, bottom panels) 
carry unpaired spins (ferromagnetic coupling between $S$=1 subsystems 
is considered there) confirming the above picture: two out of four orbitals,
\textit{a} and \textit{b}, are primarily localized on the "left-hand" side of the Ni$_2$ 
complex, while their counterparts, \textit{a'} and \textit{b'}, are localized on the 
"right-hand" side. When the Ni$_2$ complex is brought in 
contact with the Cu surface, these molecular orbitals
are transformed to resonances (\textit{a,b,a',b'}) in the spectral function 
centered around $-1.5$~eV below the Fermi-level $E_F$ (see Fig.~\ref{Fig_Ni2Spectrum}), 
while the upper Hubbard band  ($\varepsilon_d$) is placed just above $E_F$.   
Furthermore, since the wave functions \textit{a} and \textit{a'} 
involve larger contributions from $\pi$ orbitals of the central bpym unit, 
the corresponding resonances \textit{a} and \textit{a'} are much broader
than the \textit{b} and \textit{b'} ones.

Assuming that the weak AF coupling between the
two subsystems is below the Kondo-temperature, 
$J_\mathrm{AF} = (E_\mathrm{AF}-E_\mathrm{F})/2 \simeq 12\mathrm{K} \le T_\mathrm{K}$
(otherwise the AF singlet ground state would be incompatible with the 
Kondo-effect observed experimentally), 
each subsystem will undergo Kondo-screening independently below the Kondo-temperature.
We note that due to inversion center, 
each spin has its own conduction channel \cite{Jayaprakash1981}.
Using equation  (\ref{eqTK}) (limit $U \gg \varepsilon_d$), 
and parameters read from the computed spectral function (see Fig.~\ref{Fig_Ni2Spectrum}),
\begin{eqnarray}
\nonumber
& & 2\Gamma^{*} = \Gamma_{a,a'} \simeq 0.1~\mathrm{eV} \gg \Gamma_{b,b'}, \\
\nonumber
\varepsilon_d  &  \sim  &  0.06 \div 0.125~\mathrm{eV},\quad 
U \simeq 1.5~\mathrm{eV} \gg \varepsilon_d,
\end{eqnarray}
we estimate $T_\mathrm{K} \sim 0.4 \div 20~\mathrm{K}$, where 
the upper limit is above $J_\mathrm{AF}$ and is in the range of
experimental values.
\bigskip

\begin{figure}[t]
\begin{center}
\includegraphics[width=0.95\linewidth]{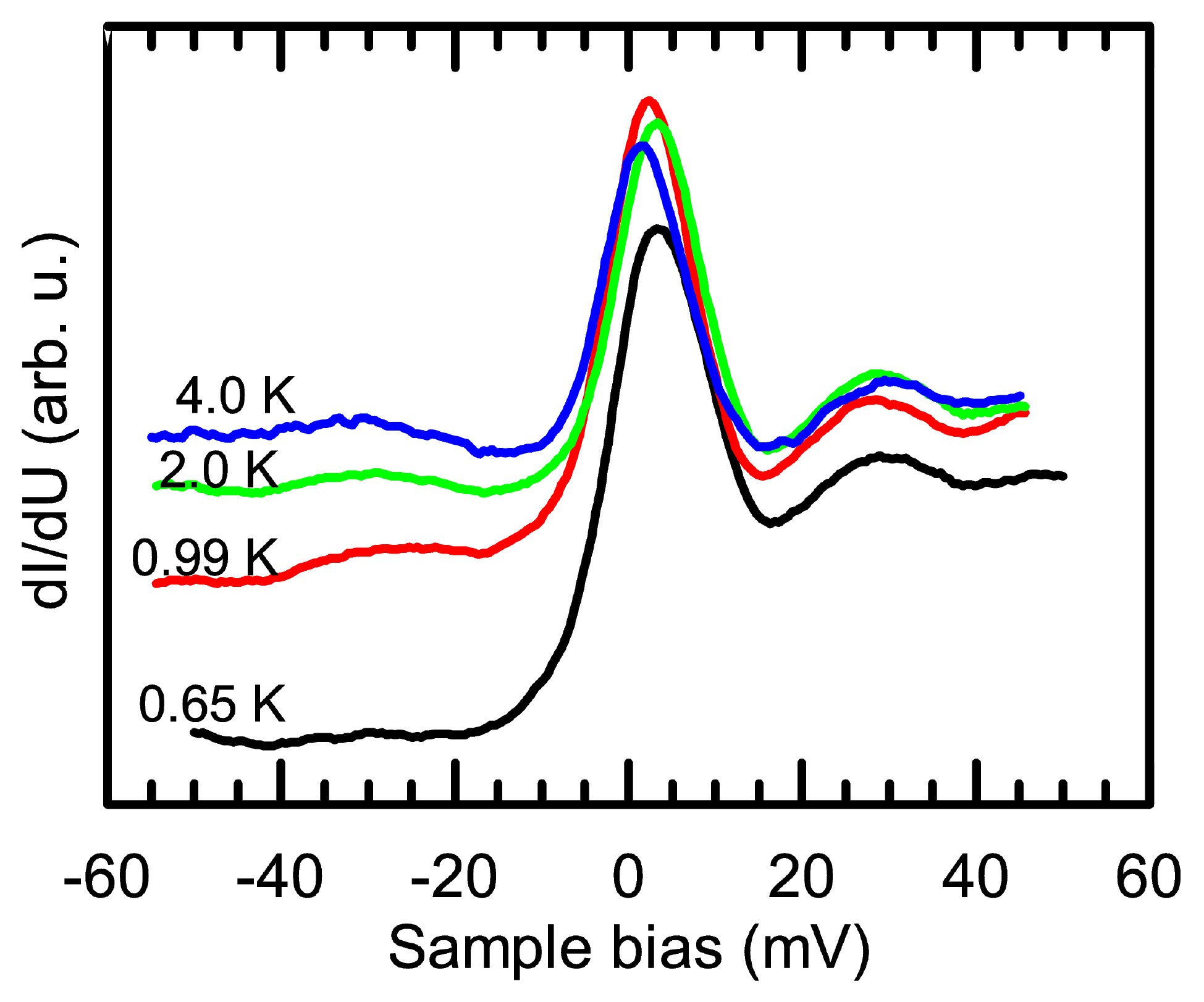}
\caption{\label{Exp_Fig4}
Differential conductance $dI/dU$ of Ni$_2$-${\beta}$ measured 
at different temperatures. Sidebands at 30~mV are
surrounding the Kondo-resonance at zero bias. 
Feedback conditions: $U=50$~mV, $I=1$~nA, 2~mV modulation.
}
\end{center}
\end{figure}

\subsection{Satellites accompanying the Kondo-resonance}

Yet another intriguing experimental observation, found in the 
differential conductance $dI/dU$ of Ni$_2$-$\beta$, 
are two satellites at $\pm 30$~mV (Fig.~\ref{Exp_Fig4}). 
We discuss further four possible hypotheses,
which might explain their origin. These are: 
(a) parallel conduction through frontier molecular orbitals (LUMOs);
(b) low-energy magnetic excitations from the singlet ground state
within the binuclear complex; (c) zero-field splitting of the
triplet state of Ni$^{2+}$ ion; and (d) low-energy
vibrational excitations of the complex and associated with that
phonon-assisted Kondo-effect. 
According to the analysis presented below, 
three hypotheses (a), (b) and (c) are likely to 
be ruled out in favor of the hypothesis (d).

\begin{figure}[t]
\begin{center}
\includegraphics[width=.95\linewidth]{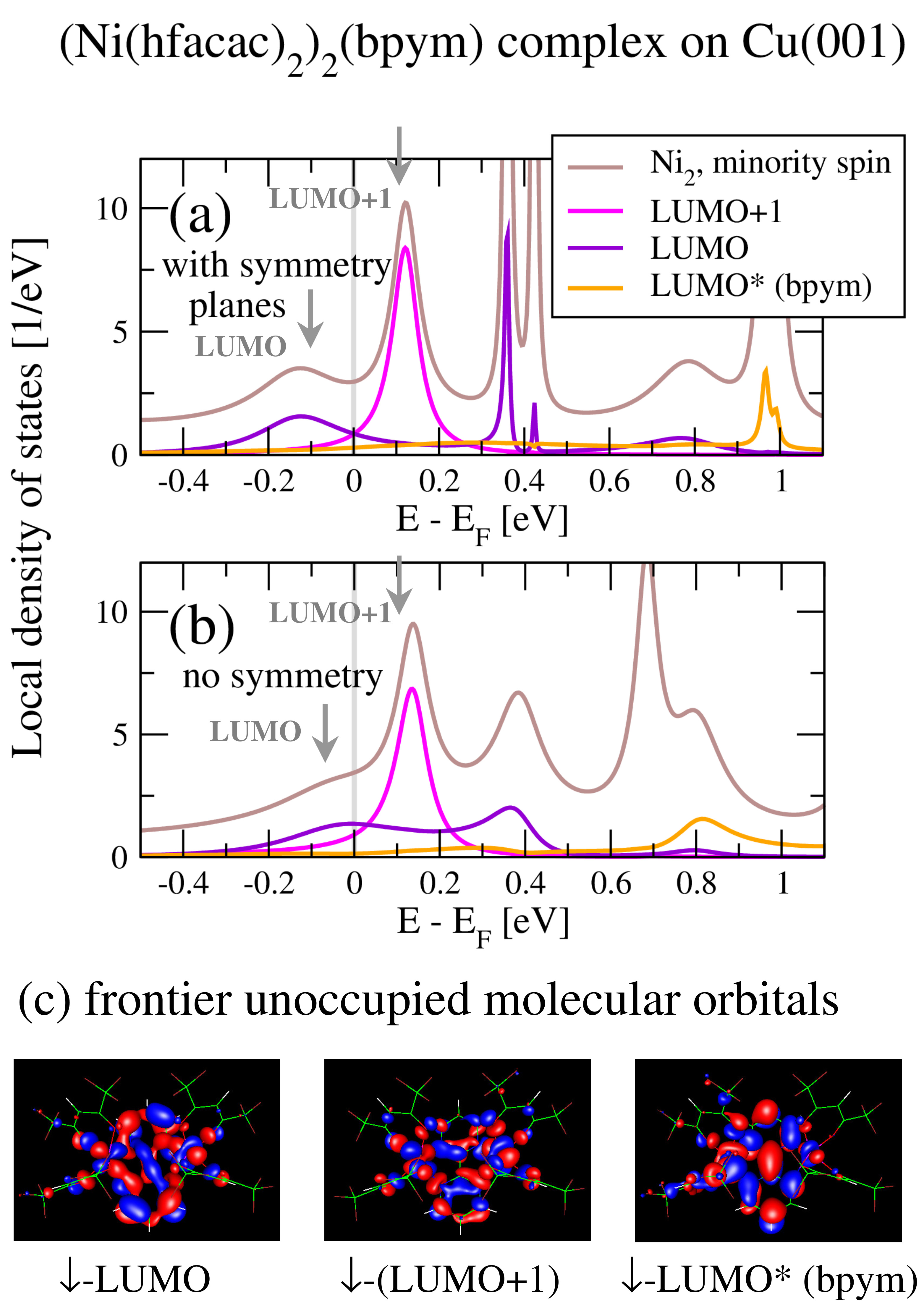}
\caption{\label{Fig_Ni2xSpectrum}
(a) and (b) Kohn-Sham spectral function of Ni$_2$ complex 
at Cu surface in the minority spin channel; panel (a): Ni$_2$ complex with
local $C_{2v}$ symmetry; panel (b): Ni$_2$ complex relaxed at surface
without symmetry constrains. Lower plot (c) shows frontier (minority-spin)
unoccupied molecular orbitals.
}
\end{center}
\end{figure}

\textit{Hypothesis (a)}. 
The first plausible suggestion is that satellites in $dI/dU$ around the Kondo-resonance 
may be attributed to the parallel conduction through the 
{\small LUMO} and {\small LUMO}+1 of a Ni$_2$ complex, 
as summarized in Fig.~\ref{Fig_Ni2xSpectrum}.
A distorted gas-phase Ni$_2$ complex has almost degenerate {\small LUMO} and {\small LUMO}+1 levels 
(e.g., in the minority spin channel, if Ni spins are coupled ferromagnetically), 
with splitting $\Delta \simeq 0.08$~eV comparable with the required energy scale $2\,\delta E \simeq 0.06$~eV.
A pair of LUMO and LUMO+1 wave functions (see Fig.~\ref{Fig_Ni2xSpectrum}c) involve 
$d$-orbitals of two Ni$^{2+}$ centers hybridized via $\pi$-states of the central 
bpym-unit. When the Ni$_2$ complex is deposited on a Cu surface (Fig.~\ref{Fig_Ni2STM}), the molecular orbitals
hybridize with the substrate states in different ways,\cite{lumos} 
as seen from Figs.~\ref{Fig_Ni2xSpectrum}a,b.
Two "satellites" are seen in the DFT spectral function $A(E)$:
below $E_F$ (a "shoulder" of the partially occupied {\small LUMO} state) 
and above $E_F$ (mainly, {\small LUMO}+1 resonance). \cite{remarkkondo}
However, these "satellites" are placed at energies 
around $\pm 100$~meV \textit{vs} the Fermi-level, 
i.e.\,above the required energy $\delta E \simeq 30$~meV.
Their positions in the spectral function $A(E)$
are furthermore sensitive to the variations in the absorption 
geometry (cf. plots a and b in Fig.~\ref{Fig_Ni2xSpectrum}), 
and it is unlikely that the found "satellites" are always 
expected at equidistant points with respect to the Fermi-level. 

\textit{Hypothesis (b)}. 
Following the data presented in Table~I, also 
magnetic excitations are unlikely to be the cause.
A transition from AF to F coupling of $S$=1 spins happens at the energy scale $\sim 2$~meV,
too small to explain peaks at $\pm 30$~meV.
Furthermore, the AF singlet ground state is not compatible with the observed Kondo-effect. 
Furthermore, breaking Hund's rule and 
flipping a 1/2 spin at the Ni$^{2+}$ center is too expensive 
($\sim 0.5$~eV, see Table~I).

\textit{Hypothesis (c)} relies on zero-field spitting of a triplet $S$=1 state 
(we refer here to one of the magnetically almost isolated 
subsystem --- a Ni(hfacac)$_2$ unit with a single Ni$^{2+}$ center). 
Because of spin-orbit interaction, 
the triplet splits into a (quasi-degenerate) doublet and a singlet, 
with the lowest energy state to be depended on the sign of the 
largest anisotropy constant $D$ in the
zero-field spin Hamiltonian, 
$H_{ZF} = D[S_z^2 - S(S+1)/3] + E(S_x^2 - S_y^2)$. 
An experimental evidence exists, Ref.~\onlinecite{CoKondo}, 
that for example under a strong distortion of the octahedral symmetry 
splitting can be of few meV. However, if 
$D>0$, the ground state is singlet, while the doublet is accessible at non-zero
bias voltages only corresponding to a split Kondo-resonance, a situation 
that is not observed experimentally. Contrary, if $D<0$, the ground state is (almost) 
degenerate, provided $|E| \ll |D|$. At zero bias-voltage
a fluctuation from $S_z=1$ to $S_z=-1$ is not easily possible, since 
$\Delta S_z = 2$ is not compatible with a spin-flip event by a single substrate 
electron with $\Delta m = 1$. 
In this situation, only a scattering event of second order involving 
two electrons can cause a Kondo-effect, which would result in extremely low Kondo-temperatures.

\textit{Hypothesis (d)}. 
Under applied bias voltage, a "hot" electron tunneling from the STM-tip
(or from the surface) through a molecule
may release its energy via emitting a phonon  
before participating in the spin-flip scattering processes.  
Such an inelastic Kondo-effect would manifest 
as a "copy" of the Kondo-resonance in $dI/dU$, 
which is, however, shifted from zero bias to the energy of 
molecular vibration \cite{inelastic-Kondo}.

To verify this idea, we have performed quantum-chemistry 
calculations of the electron-phonon coupling matrix elements
for the Ni$_2$ complex.
To simplify our analysis, we have considered a distorted molecular 
conformation of Ni$_2$ with imposed $C_{2v}$ symmetry that
closely resembles the atomic structure of the complex pre-optimized 
nearby a Cu(001) surface (see Suppl.~Figs.~1b and 4a in Supplementary Information\cite{suppinfo}). 

\begin{figure}[t]
\begin{center}
\includegraphics[width=0.95\linewidth]{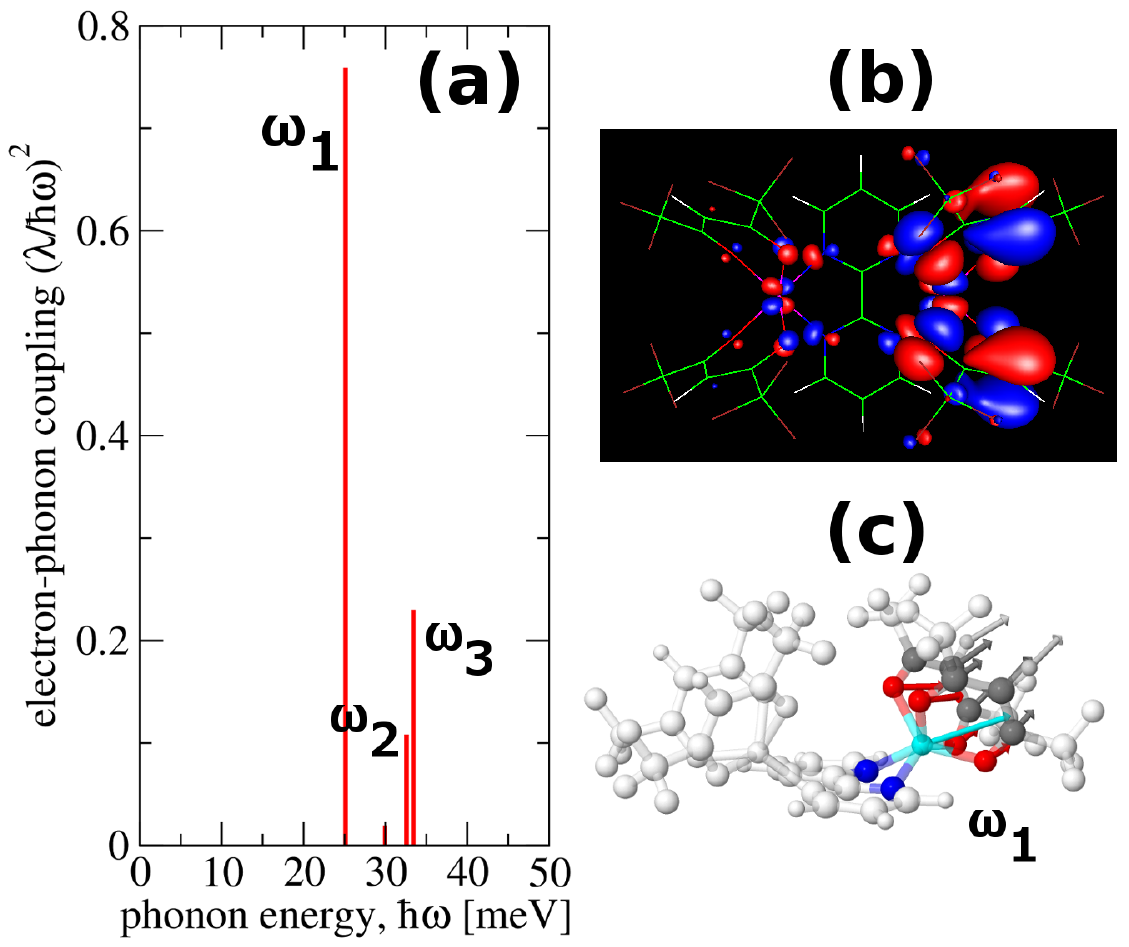}
\caption{\label{Fig_Phonons}
Dimensionless electron-phonon (\textit{el-ph}) coupling constants, 
plot (a), between the "Kondo-active" molecular orbital 
$\psi_K$ shown in plot (b) 
and low-energy vibrational eigenmodes, which are forced to be 
localized in the vicinity of Ni$^{2+}$ ion at the right-hand side
of the molecule (see text for details).
Plot (c) is a visual representation of the mode (atomic displacements 
are scaled by $\times$10) with energy $
\hbar\omega_1 = 25.1$~meV exhibiting the largest 
\textit{el-ph} coupling constant $\approx 0.75$. 
}
\end{center}
\end{figure}

To first order, molecular 
vibrations of two Ni(hfacac)$_2$ units can be considered independently, 
since these units are attached to the bpym-moiety, which is 
strongly bound to the Cu surface. Therefore, we consider one of  
the Ni(hfacac)$_2$ units and define an "active" subspace 
limited to 15 atoms including one Ni$^{2+}$ ion and surrounding atoms 
in its vicinity (see Fig.~\ref{Fig_Phonons}). There "colored" atoms are 
allowed to vibrate, while all other "reservoir" atoms shown in gray 
are assumed (as an approximation) to have infinite masses: either because 
some of them are heavy trifluoromethyl CF$_3$ groups, or because 
other atoms are bound to the surface. Such an approximation provide us with 
a set of 45 well-defined vibrational frequencies~$\omega_{\mu}$.

Further, we consider the wave function $\psi_{K}$ 
(Fig.~\ref{Fig_Phonons}b), which dominates in  
the scattering channel responsible for the 
Kondo-effect (essentially, $\psi_{K}$ is one of those
single-occupied molecular orbitals 
depicted as $a$ and $a'$ in Fig.~\ref{Fig_Ni2xSpectrum}, 
which strongly hybridize with the Cu surface) \cite{kondo-orbital}.
We have computed electron-phonon coupling matrix elements 
$\lambda^{\mu}$, involving the orbital  $\psi_{K}$
and molecular vibrations, which are localized
within the same "active" subspace (for computational details, 
see Supplementary Information,\cite{suppinfo} Sec.~III).

Our results are presented in Fig.~\ref{Fig_Phonons} in which 
we show the dimensionless electron-phonon 
coupling constants $g = (\lambda^{\mu}/\hbar\omega_{\mu})^2$ limited to
the low-energy excitations. We observe only three eigenmodes 
(25.1~meV, 32.6 meV and 33.4 meV) with non-zero coupling 
constants and frequencies in the proximity of $\delta E \simeq 30$~meV. 
Furthermore, we show in the Supplementary Information,\cite{suppinfo} Sec.~III, 
that the eigenmodes involve vibrations of Ni$^{2+}$ ions. 
Their energies will be renormalized when coupled to a continuum of 
vibrational modes of the macroscopic system, including  
remaining functional groups of the molecular complex and the Cu surface. 
However, our additional calculations show that, for example, interaction between two 
Ni-subsystems introduces a moderate splitting of the frequencies only, 
around $\sim 0.5$~meV. Thus, we argue that three vibrational 
eigenmodes may rationalize the observation of satellites in $dI/dU$ 
as a signature of the phonon-assisted Kondo-effect.

\section{Conclusion}

To summarize, low temperature STS measurements on binuclear metal-organic
complexes, "Ni$_2$" and "Mn$_2$", deposited on a Cu(001) surface revealed that 
the systems undergo the Kondo-effect with the Kondo-resonances located nearby transition metal atoms. 
The relatively large Kondo-temperatures, of the order of $\sim$~10~K, were 
found to depend on the adsorption-type. The situation was intriguing here, because 
the synthesized molecules do not have predefined anchoring groups, which could be 
responsible for a formation of the chemical bond with the Cu surface.\cite{Ruben2013} 

We rationalized experimental observations by performing
extensive density functional theory calculations. 
We searched for the adsorption geometries, where 
molecules are chemisorbed on the surfaces. 
In case of "Ni$_2$", our simulations 
show that some STM images (Ni$_2$-$\beta$)
can be attributed to a distorted "Ni$_2$" complex with partially weakened 
internal chemical bonds, while other STM images (Ni$_2$-$\alpha$) may be 
interpreted as arising from molecular fragmentation. 
In both cases, our calculations suggest a picture of the 
underscreened ($S$=1)-type Kondo-effect emerging from the open 
3$d$ shells of the individual Ni$^{2+}$ ions.
Furthermore, theoretical analysis points out that
the satellites in the STS spectra
observed nearby the zero-bias resonance 
are likely a signature of the low-energy vibrational excitations of the
"Ni$_2$" complex and associated with that phonon-assisted Kondo-effect. 

In broader terms, binuclear complexes present 
an excellent playground for studying fundamental aspects
of magnetic two-impurity (or double quantum dot) systems. By functionalizing the
bridging unit it could be possible to enhance the super-exchange interaction between two
centers, thus allowing to access different regions of the phase diagram of the double
impurity model \cite{Jayaprakash1981,Hofstetter,Bork2011}. On the applied level, 
understanding interaction between the spins residing on different functional 
units is vital to quantum information storage and processing with molecules assembled
on surfaces. Our work provides an important step in this direction and identifies
further challenges. Apart from the appealing enhancement of the super-exchange, 
the substitution of Ising-like spins (such as Tb) may offer means
to study a "double molecular magnet".  As demonstrated here, first-principles 
calculations will play an indispensable role in this effort.
\bigskip
\smallskip

\section*{Acknowledgments}
We thank Karin Fink for fruitful discussions
and Marius B\"{u}rkle for providing us with a development
version of the {\small{TURBOMOLE}} package.
A.B.\ acknowledges support of DFG through the research 
grant BA 4265/2-1. W.W.\ and M.R.\  
acknowledge support from the Baden-W{\"u}rttemberg 
Stiftung in the framework of the Kompetenznetz f{\"u}r Funktionale Nanostrukturen (KFN).



\begin{thebibliography}{99}
\bibitem{Hewson1}
The formula for the Kondo temperature reads as: 
\[
k_B T_\mathrm{K} \simeq 0.41 U \sqrt{\frac{\Gamma}{4U}}\ e^{-\frac{\pi(U-\varepsilon_d)\varepsilon_d}{U\Gamma}},  
\]
where $-\varepsilon_d < 0$ is a position of the occupied resonance level measured \textit{vs} Fermi-energy,
$\Gamma$ is the resonance width and $U$ is Coulomb on-site interaction.
\bibitem{lumos}
Our calculations indicate that 
a fractional charge of $\approx 0.5e$ leaks out from the surface to the molecule, 
populating partially {\small LUMO} of Ni$_2$ complex and {\small LUMO}* of 
the bpym-moiety (see Fig.~\ref{Fig_Ni2xSpectrum}). Also, a magnetic moment 
carried by the complex is slightly reduced, from $4.0\mu_B$ 
(gas-phase molecule) down to $3.3\mu_B$.
\bibitem{remarkkondo}
We remind that DFT with a quasi-local XC-functional does not 
allow to recover a Kondo-resonance at the Fermi-level.
\bibitem{kondo-orbital}
More specifically, "Kondo-active" molecular orbitals (MOs)
depicted as $a$ and $a'$ in Fig.~\ref{Fig_Ni2xSpectrum}, 
which are strongly coupled to the Cu surface, are well 
approximated by symmetric and antisymmetric combinations
$\psi^{\pm}_{K} = \frac{1}{\sqrt{2}}(\phi^{\uparrow}_{b_1} \pm \phi^{\uparrow}_{a_2})$ 
of the single-occupied MOs of the distorted Ni$_2$ complex 
with $C_{2v}$ symmetry. Here $b_1$ and $a_2$ refer to irreducible representations of 
the $C_{2v}$ group, with $b_1$ being even and $a_2$ being odd with 
respect left-right reflection $\sigma_v$.
\end{thebibliography}
\end{document}